# Large anomalous Nernst effect across the magneto-structural transition in bulk Ni-Co-Mn-Sn full Heusler alloy

Arup Ghosh[1], Avirup De[1] and Sunil Nair[1,2]

[1]*Department of Physics, Indian Institute of Science Education and Research, Dr. Homi Bhabha Road, Pune 411008, India.*

[2]*Centre for Energy Science, Indian Institute of Science Education and Research, Dr. Homi Bhabha Road, Pune 411008, India.*

## Abstract

We report on the observation of temperature and field dependent anomalous Nernst effect (ANE) in Ni-rich bulk Ni-Co-Mn-Sn full Heusler alloy. A large change in the transverse Nernst coefficient ($N$) is obtained across the first order magnetostructural transition from a tetragonal martensite to a cubic austenite phase. The saturation of ANE and magnetic data appear to depend largely on the magnetic anisotropy of the device. Such change in the Nernst co-efficient may prove to be useful for switching applications controlled by temperature and magnetic field changes.

## I. Introduction

Effects of thermal gradients (*ΔT*) and magnetic fields on magnetic materials have significantly advanced the field of spincaloritronics in the past few years.[1–11] Anomalous Nernst effect (ANE) has recently received a significant amount of attention as an alternative means of evaluating thermoelectric devices for a part of future applications.[2,11–13] In the Nernst Effect, thermally excited charge carriers get deflected in the transverse direction of the magnetization due to the Lorentz force while moving to the colder side. Although, the conversion efficiency of heat to transverse thermopower is still small, one can use the waste heat and convert it for spincaloritronic applications. The origin of ANE and the anomalous Hall Effect (AHE) are very similar.[12,14] An applied *ΔT* moves the carries in ANE whereas; an electrical bias is necessary in case of AHE to move the carriers in the direction perpendicular to the magnetization of the device. The longitudinal Seebeck coefficient ($S_{xx}$) can be described using the Mott's relation[15,16]

$$S_{xx} = \frac{\pi^2 k_B^2 T}{3e\sigma}\left(\frac{\partial \sigma_{xx}}{\partial E}\right)_{E_F} \quad (1)$$

where $k_B$, $T$, $e$, $\sigma$, $E$ and $E_F$ are the Boltzmann constant, temperature, electric charge, electrical conductivity, energy and Fermi energy respectively. In the presence of magnetic field, the transverse contributions appears as $\boldsymbol{V_{ANE}} = N\ 4\pi\ \boldsymbol{M} \times \boldsymbol{\Delta T} = S_{xy}\ \boldsymbol{m} \times \boldsymbol{\Delta T} = \theta_{ANE}\ S_{xx}\ \boldsymbol{m} \times \boldsymbol{\Delta T}$, where $V_{ANE}$, $N$, $M$, $\Delta T$, $S_{xy}$, $m$ and $\theta_{ANE}$ are the measured ANE voltage, Nernst coefficient, magnetization, thermal gradient, transverse Seebeck coefficient, unit vector of the magnetization and the ANE angle respectively.[13] As in the case of the AHE, the origin of the ANE can be traced to contributions from asymmetric skew scattering,[16] side jump mechanism and the intrinsic Berry curvature.[17] $S_{xy}$ can be written as[16]

$$S_{xy} = \rho\left(\alpha_{xy} - S_{xx}\sigma_{xy}\right) \tag{2}$$

where, $\rho$ is the bulk resistivity, $\sigma_{xy}$ is the transverse electrical conductivity and $\alpha_{xy}$ is the transverse thermal conductivity which can also be derived using the Mott's relation as[12,18]

$$\alpha_{xy} = \frac{\pi^2 k_B^2 T}{3e}\left(\frac{\partial \sigma_{xy}}{\partial E}\right)_{E_F} \tag{3}$$

Till date, the ANE has been reported in a few systems like metallic and semiconducting thin films, their multilayered devices and antiferromagnets.[2,11–13,19–22] Since half Heusler alloys are recognized as one of the most promising spintronic materials due to its large spin polarization, a number of reports exists on half Heuslers showing spintronic, spin Seebeck and anomalous Hall signals.[8,10,23,24]

On the other hand, Full Heusler alloys are well known for showing large magnetocaloric effect (MCE)[25–35] across their first order magneto-structural transition (FOMST). Ni-rich stoichiometric $Ni_2MnSn$ full Heusler alloys have a cubic ($L2_1$) austenite parent phase characterized by four interpenetrating face centered cubic (fcc) sublattices (Fig. 1(a)).[30,35] Its off-stoichiometric compositions can undergo a FOMST to a tetragonal martensite structure on cooling (Fig. 1(b)), and this transition involves a significant change in the magnetic, electrical and thermal transport properties.[26,28,34,36,37] Typically, the magnetization increases sharply with the increase in temperature across the transition due to the increase in ferromagnetic interactions between the inter-site Mn atoms. The resistivity has an opposite dependence and decreases drastically on increasing the temperature across the FOMST. The ANE is expected to be proportional to the magnetization and inversely proportional to the resistivity, and thus, one can expect a giant enhancement in $V_{ANE}$ across the FOMST. The structural transition of full Heusler alloys can be tuned by temperature, field and hydrostatic pressure, and thus, any change in these aforementioned parameters across the FOMST would be expected to significantly alter the ANE properties of full Heusler alloys. This could provide an avenue for spin-caloritronic applications for low current to high current fast switching and vice versa.

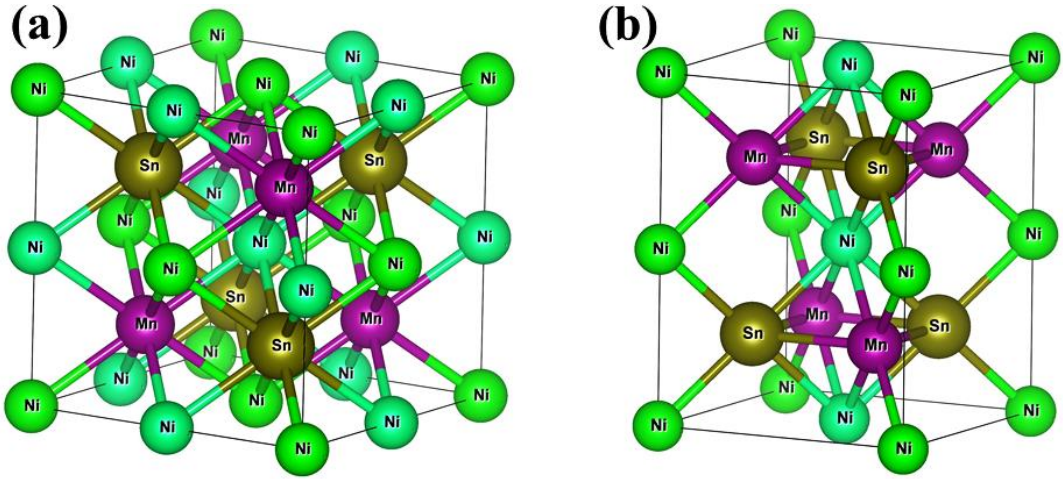

FIG. 1. Unit cell of (a) cubic $L2_1$ and (b) tetragonal full Heusler alloy.

Here, we report on the ANE as measured in a bulk Ni-rich $Ni_{46.5}Co_2Mn_{37}Sn_{14.5}$ Heusler alloy as a function of temperature, $\Delta T$ and magnetic field. A dramatic increase of the ANE is observed across the FOMST in this system. A systematic comparison has also been carried out with the magnetic measurements. Interestingly, the ANE is found to show a preference of easy direction which mimics the anisotropy exhibited by the magnetization of the device.

**III. Experimental details**

The details of the preparation and characterization of $Ni_{46.5}Co_2Mn_{37}Sn_{14.5}$ (NCMS) Heusler alloy is already mentioned elsewhere.[33] The sample was cut using an IsoMet Buehler low speed saw to form a rectangular shape of dimension 9×4.5×1 $mm^3$. The magnetic properties of the sample were measured using a magnetic property measurement system (MPMS, Quantum

Design). The resistivity and ANE were measured using a homemade setup for temperature and field dependent spin-caloritronic measurements. The setup comprises of an Advanced Research System (ARS) made closed cycle refrigerator (CCR) which can be operated from ~10 K to 320 K, a Lakeshore 340 temperature controller, a Keithley nano-voltmeter (model 2812A), an Agilent source meter (model 2400) and an electromagnet with an upper field limit of 2 kOe.

**III. Results and discussion**

Fig. 2 shows the temperature dependence of field cooled cooling (FCC) and field cooled warming (FCW) magnetization in presence of 1 kOe magnetic field in the temperatures between 2 K and 300 K. The same figure also depicts the zero field temperature dependent resistivity of the NCMS. The sample undergoes a phase transition near 225 K which is the FOMST of this alloy family. Initially, the ZFC curve shows an increase in magnetization with increase in temperature. The low temperature tetragonal phase has mixed ferro-antiferro interaction and therefore, shows low magnetization. Across the structural transition, the sample's magnetization increases dramatically. The thermal hysteresis between cooling and warming curves confirms the first order nature of the phase transition. Such a large change in the magnetization across this transition gives rise to a large MCE. When the structure changes from tetragonal to cubic, the inter-site Mn-Mn separation increases which in turn enhances the ferromagnetic interaction in NCMS. As a result, the magnetization increases on traversing the FOMST from the low temperature to the high temperature side. In the case of resistivity, both the structural phases show metallic behavior, The resistance decreases across the martensitic transition due to the change of phase from lower symmetric tetragonal to a highly symmetric cubic one. As the ANE is proportional to the magnetization of a magnetic conductor, one can expect a huge change in the ANE voltage ($V_{ANE}$) across the structural transition of NCMS.

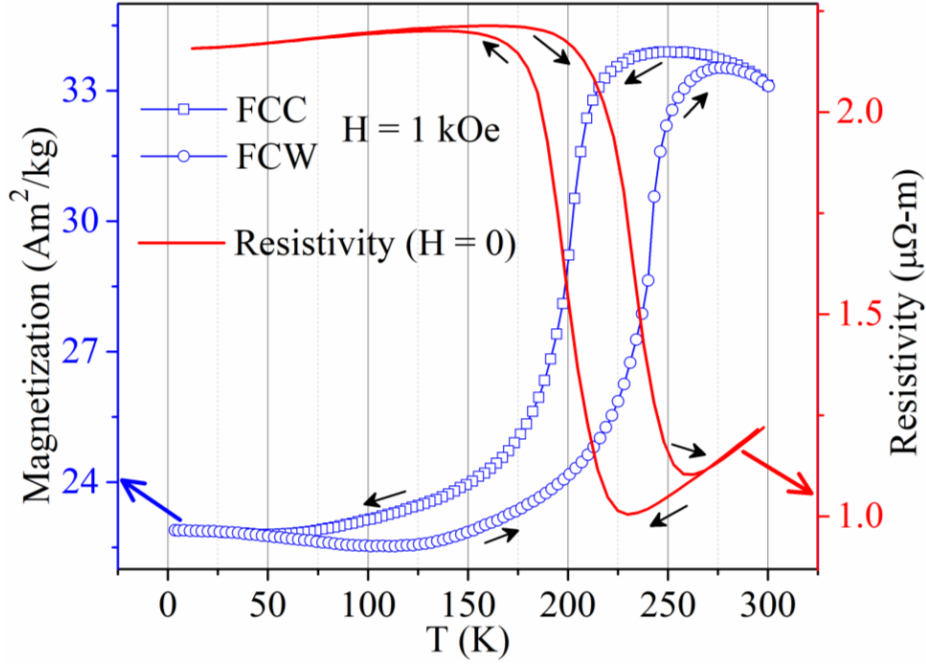

FIG. 2. Temperature dependence of magnetization (M-T curves) under 1 kOe field and zero field resistivity for NCMS.

Fig. 3(a) gives a schematic representation of the sample preparation process; arc-melting, cutting and final form of device. The temperature dependent normalized $V_{ANE}$ is plotted in Fig. 3(b) for ZFC warming and cooling modes. The FCC and FCW temperature dependent magnetization (magnetization axis is plotted in reverse order) are also plotted together with $V_{ANE}$ to compare and match the transition. A schematic diagram of the device showing the direction of measurement of ANE voltages is given in the inset of Fig. 3(b). Here, the $\varDelta T$ is applied along the width ($L_Z$) and the magnetic field is kept in plane along the breadth ($L_Y$) of NCMS sample. Therefore, the induced $V_{ANE}$ appears along the length ($L_X$) of the device. The signal is normalized using the equation

$$V_{ANE}^{Normalized} = \frac{\left(V_{ANE}/L_X\right)}{R_X\left(\Delta T/L_Z\right)} \qquad (4)$$

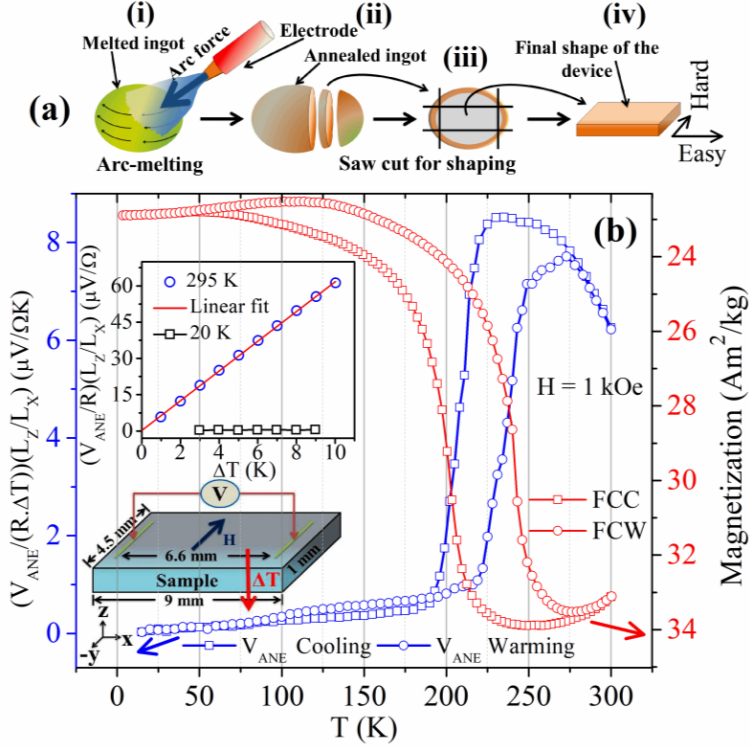

FIG. 3. (a) Schematic representation of the processing of bulk Heusler alloy device from arc-melting. (b) Temperature dependent normalized anomalous Nernst voltage and resistivity of the NCMS device along the length configuration. Inset: Thermal gradient dependent ANE at 20 K and 295 K.

The FOMST is clearly observable in the ANE measurement and it strictly follows the magnetic measurement. At the lower temperatures, $V_{ANE}$ decreases with the decrease in temperature due to the reduction in the magnon population which in turn suppresses the scattering probability of electrons by the magnons. The temperature dependence of the measured ANE follows the well-known Mott relation where skew scattering and the side jump mechanisms contributes significantly.[9,12,15] Chuang et al.[16] substituted $\sigma_{xy}$ by the power-law scaling of $\rho_{xy} = \lambda M \rho^n$, where $\lambda$ represents the spin-orbit coupling strength, giving

$$\frac{N}{\rho^{n-1}} = \frac{\pi^2 k_B^2 T}{3e} \lambda' - (n-1)\lambda S_{xx} \qquad (5)$$

Here, n = 2 value corresponds to the intrinsic Berry curvature and side jump mechanism, whereas the skew scattering have n = 1 where the second term in Eq. (2) vanishes. Their results on permalloy, Fe, Co and Ni thin films have shown that n becomes 1 for larger $L_Z$ (> 20 nm). In the present study, $L_Z$ ~ 1 mm and thus the skew scattering mechanism would be expected to be the predominant contributor in the ANE of the NCMS system.

Across the FOMST, the increase in magnetization of the NCMS enhances the ANE. Moreover, resistance of the device drops significantly (more than 50%) during the structural transition which gives rise to a change in order of magnitude for normalized ANE. Furthermore, we have estimated $N$ of our sample just below and after the structural transition using the following equation

$$N = \frac{1}{\mu_0 H}\left(\frac{V_{ANE}/L_X}{\Delta T/L_Z}\right) \quad (6)$$

where, $\mu_0 H$ is the applied magnetic flux density. $N$ increases from ~0.03 µVK$^{-1}$T$^{-1}$ to ~0.13 µVK$^{-1}$T$^{-1}$ during the structural transition from martensite to austenite. The change in $N$ across the magneto-structural transition is one order of magnitude. This can be useful for switching purposes where a spintronic circuit can be turned to high current state from a low current state and vice versa by changing the temperature. The same can also be achieved by the use of a magnetic field as a tuning parameter, since these alloys are known to exhibit field induced phase transitions from the martensitic to austenite phases.[38,39]

The inset of Fig. 3(b) shows the thermal gradient dependence of normalized $V_{ANE}$ at 295 K and 20 K under 1 kOe magnetic field, and the observed linearity of $V_{ANE}$ with $\Delta T$ is consistent with the equation $V_{ANE} \sim \Delta T \times M$. At 20 K, the ANE signal has drastically dropped due to the decreased magnon-electron scattering and the increased coercivity of the NCMS which does not allow us to drive the sample to saturation within our given field limit.

The ZFC isothermal magnetic hysteresis loop at 75 K is plotted in Fig. 4(a). Fig. 4(b) represents the ZFC field dependent normalized $V_{ANE}$ for NCMS at 75 K. In the martensite phase, the saturation field is high due to large magneto-crystalline anisotropy and ferrimagnetic nature of the NCMS. Thus, the loops do not saturate completely. The ZFC hysteresis loops look symmetric with respect to the magnetic field axis, which confirms the absence of any EB in low field ($\leq 2$ kOe).

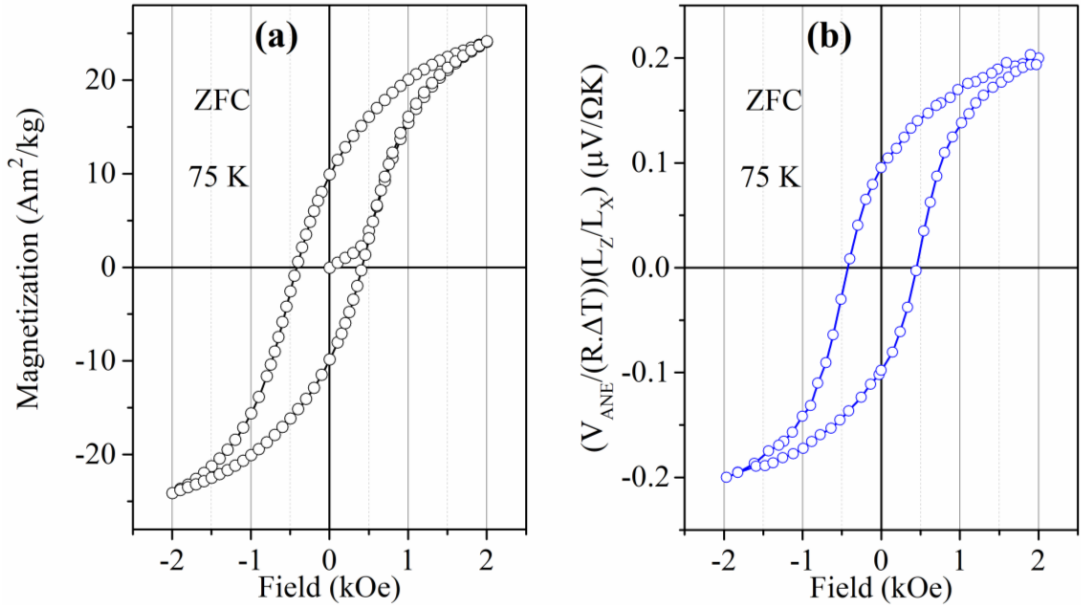

FIG. 4. (a) Zero field cooled (ZFC) magnetic hysteresis loop and (b) ZFC normalized ANE signal measured as a function of magnetic field at 75 K.

Fig. 5(a) and 5(b) show the isothermal field dependence of $V_{ANE}$ and magnetization of the device measured at 300 K. For $V_{ANE}$, the lower and upper surface of the device was kept at 305 K and 295 K keeping the $\Delta T$ constant at 10 K. The magnetic field dependence of both the ANE and magnetization exhibit similar trend. These measurements were performed in two different configurations; along the length (AL) and along the breadth (AB) as depicted in Fig. 5(c) and 5(d) respectively. Interestingly, the magnitude and $H_S$ differ significantly for AL and AB configurations. Magnetic anisotropy in systems like the one which we are investigating could arise from contributions from the magnetocrystalline anisotropy, uniaxial anisotropy, stress anisotropy, and shape anisotropy.[40] At room temperature, the crystallographic structure is cubic, and hence the uniaxial contribution would be expected to be absent. Since the devices being investigated are polycrystalline in nature, the contribution from the magnetocrystalline anisotropy would also average out to a large extent. However, the strain fields frozen in the ingot during the recrystallization process could contribute to the effective magnetic anisotropy. Prior density-functional theory (DFT) calculations have indicated that within the cubic phase, the $Ni_2MnSn$ system has the highest elastic anisotropy factor amongst all the members of the extended Ni2Mn*M* (with *M* = Al, Ga or Sn) family.[41] A schematic diagram is given in Fig 3(a) which represents the sample preparation technique using arc-melting and subsequent processing of the samples for device preparation. We have prepared the device by cutting the ingot as shown in Fig. 3(a). In addition, the shape anisotropy arising due to geometry dependent demagnetization factors could also play an important role. The final device is observed to have its easy magnetization direction along the length (as shown in Fig. 3(a(iv))). Our measurements show that the $V_{ANE}$ in AB configuration also saturates quickly as the magnetic field is applied along $L_X$ of the device which is the easy direction.[42,43] This can be very useful for spintronic circuits where the same device can provide two different ANE responses depending on the direction of the applied thermal gradient and the magnetic field.

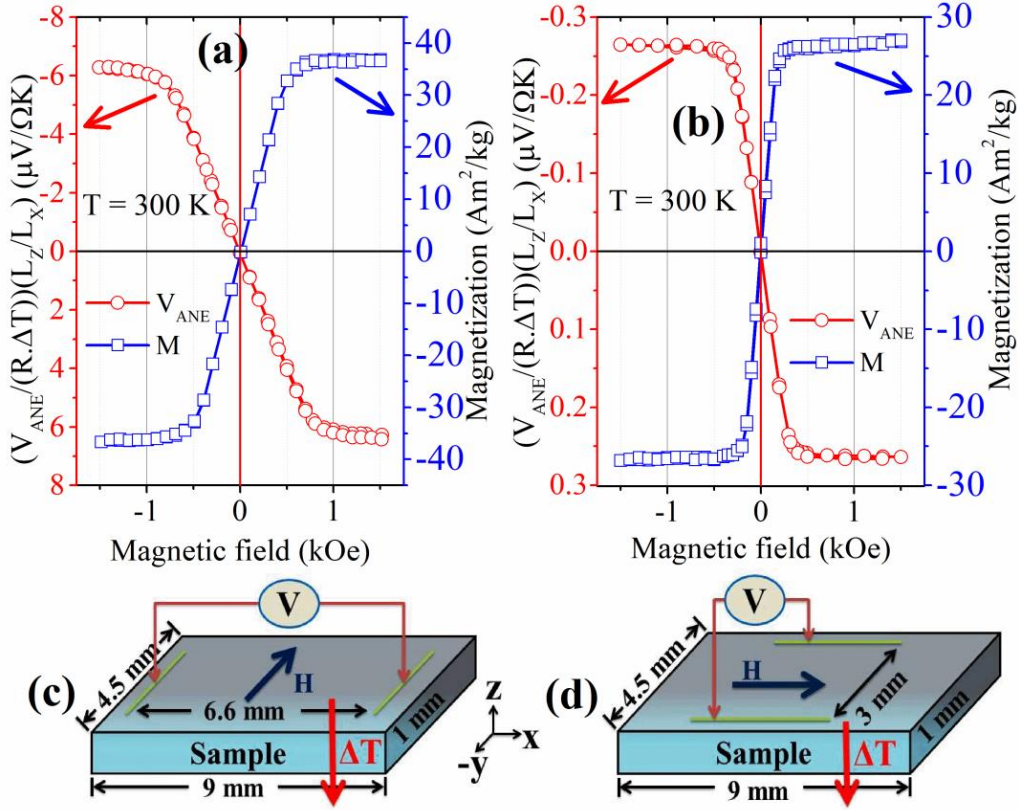

FIG. 5. Normalized anomalous Nernst signal and magnetization of the NCMS device at room temperature measured as a function of magnetic field for (a) along the length (AL) and (b) along the breadth (AB) configuration. Schematic diagram for (c) along the length (AL) and (d) along the breadth (AB) measurement configuration.

### III. Conclusion

In summary, we have systematically studied the anomalous Nernst effect in $Ni_{46.5}Co_2Mn_{37}Sn_{14.5}$ full Heusler alloy. The spincaloric measurements appear to mimic the magnetic data. The Nernst coefficient increases by one order of magnitude across the magneto-structural transition, which can be very useful for controlled switching applications by changing the temperature and magnetic field. The magnetic anisotropy of the device is seen to affects the spincaloric property significantly in different orientations.

**Acknowledgements:**

Arup Ghosh is thankful to SERB, DST, Govt. of India for providing the financial support through National Post-Doctoral Fellowship (PDF/2015/000599). SN acknowledges funding support by the Department of Science and Technology (DST, Govt. of India) under the DST Nanomission Thematic Unit Program (SR/NM/TP-13/2016).

2765 (1995).